# Large anomalous Hall effect driven by non-vanishing Berry curvature in non-collinear antiferromagnetic Mn$_3$Ge

Ajaya K. Nayak,[1,2,*] Julia Erika Fischer,[1] Yan Sun,[1] Binghai Yan,[1] Julie Karel,[1] Alexander C. Komarek,[1] Chandra Shekhar,[1] Nitesh Kumar,[1] Walter Schnelle,[1] Jürgen Kübler,[3] Claudia Felser,[1] Stuart S. P. Parkin[2,*]

[1]Max Planck Institute for Chemical Physics of Solids, Nöthnitzer Str. 40, 01187 Dresden, Germany

[2]Max Planck Institute of Microstructure Physics, Weinberg 2, D-06120 Halle, Germany

[3]Institut für Festkörperphysik, Technische Universität Darmstadt, 64289 Darmstadt, Germany

**It is well established that the anomalous Hall effect that a ferromagnet displays scales with its magnetization. Therefore, an antiferromagnet that has no net magnetization should exhibit no anomalous Hall effect. Here we show that the non-collinear triangular antiferromagnet Mn$_3$Ge exhibits a large anomalous Hall effect comparable to that of ferromagnetic metals; the magnitude of the anomalous conductivity is ~500 (Ω. cm)$^{-1}$ at 2 K and ~50 (Ω.cm)$^{-1}$ at room temperature. The angular dependence of the anomalous Hall effect measurements confirm that the small residual in-plane magnetic moment has no role in the observed effect. Our theoretical calculations demonstrate that the large anomalous Hall effect in Mn$_3$Ge originates from a non-vanishing Berry curvature that arises from the chiral spin structure, and which also results in a large spin Hall effect of 1100 (ℏ/e) (Ω.cm)$^{-1}$, comparable to that of platinum. The present results pave the way to realize room temperature antiferromagnetic spintronics and spin Hall effect based data storage devices.**

In a normal conductor, the Lorentz force gives rise to the Hall effect, wherein the induced Hall voltage varies linearly with the applied magnetic field. Moreover, in a ferromagnetic material one can realize a large spontaneous Hall effect, termed the anomalous Hall effect (AHE), that scales with the magnetization (*1, 2*). Although the detailed origin of the AHE has been puzzling for many years, recent theoretical developments have provided a deeper understanding of the intrinsic contribution of the AHE in ferromagnets in terms of the Berry-phase curvature (*1, 3-6*). The Berry phase is a very useful tool for understanding the evolution of an electron's wave-function in the presence of a driving perturbation such as an electric field. The electron's motion may deviate from its usual path, wherein the electron follows a path along the direction of the applied electric field, and may pick up a transverse component of momentum even in the absence of an external magnetic field, leading to a Hall effect. This process can be understood by calculating the Berry curvature, which is an intrinsic property of the occupied electronic states in a given crystal with a certain symmetry, and only takes a non-zero value in systems where time-reversal symmetry is broken, for example, by the application of a magnetic field, or where magnetic moments are present. From symmetry analysis, the Berry curvature vanishes for conventional anti-ferromagnets with collinear moments, thereby resulting in zero AHE. However, a non-vanishing Berry-phase in antiferromagnets with a non-collinear spin-arrangement was recently predicted leading to a large anomalous Hall effect (*7-9*). In particular, Chen *et al.* have demonstrated theoretically that an anomalous Hall conductivity (AHC) larger than 200 $(\Omega.cm)^{-1}$ can be obtained in the non-collinear antiferromagnet $Mn_3Ir$ (*7*).

The Heusler compound $Mn_3Ge$ is known to crystallize in both the tetragonal and hexagonal structures depending upon the annealing temperature (*10, 11*). Hexagonal $Mn_3Ge$ exhibits a triangular antiferromagnetic (AFM) structure with an ordering temperature of ~365-400 K (*11-14*). As shown in Fig. 1A, the hexagonal unit cell of $Mn_3Ge$ consists of two layers of Mn triangles stacked along the *c*-axis. In each layer, the Mn atoms form a Kagome lattice with Ge sitting at the center of a hexagon. Neutron diffraction studies have revealed a non-collinear triangular AFM spin-arrangement of the Mn spins in which the neighboring moments are aligned in-plane at a 120-degree angle (*13, 14*). These same studies have reported that the spin triangle is free to rotate in an external magnetic field (*13, 14*). This non-collinear structure arises due to the geometrical frustration of the Mn moments that are arranged in a triangular configuration in the *a-b* plane. We have performed *ab-initio* calculations of the magnetic ground state of $Mn_3Ge$. The lowest energy spin configuration is shown in Fig. 1B and is consistent with previous calculations (*15, 16*). In addition to the 120-degree ordered AFM, a very small net in-plane moment exists due to a small tilting of the Mn moments (*13*).

We have calculated the Berry curvature and AHC in $Mn_3Ge$ using the linear-response Kubo formulism (*1*) based on our calculated magnetic structure (Fig. 1B). Interestingly, we find that the AHC exhibits nearly zero values for both *xy* ($\sigma^z_{xy}$) and *yz* ($\sigma^x_{yz}$) components and only large values for the *xz* component $\sigma^y_{xz}$ = 330 $(\Omega.cm)^{-1}$. Here $\sigma^k_{ij}$ corresponds to the AHC when current flows along the *j* direction and thereby results in a Hall voltage along the *i* direction: $\sigma^k_{ij}$ can be treated as a pseudovector pointing along the *k* direction that is perpendicular both to the *i* and *j* directions. The symmetry of the antiferromagnetic structure of $Mn_3Ge$ determines which components of AHC tensor must be zero. As shown in Fig.1, the two magnetic layers that compose the primitive unit cell of the AFM of $Mn_3Ge$ can be transformed into each other by a mirror reflection with respect to the *xz* plane, as indicated in Fig. 1, plus a translation by *c*/2 along the *c* axis. Due to the mirror symmetry, the components of the $\sigma^k_{ij}$ vanish if they align parallel to the mirror plane. This is the case for $\sigma^z_{xy}$ and $\sigma^x_{yz}$, which are expected to be exactly zero. However, the aforementioned residual net in-plane moment acts as a perturbation of the mirror symmetry. As a consequence, $\sigma^z_{xy}$ and $\sigma^x_{yz}$ can pick up nonzero, but tiny values. By contrast, the component of $\sigma^k_{ij}$ perpendicular to the mirror plane, namely $\sigma^y_{xz}$, is nonzero. We show the calculated values of $\sigma^y_{xz}$ in the *xz* plane in Fig. 1D. The distribution is highly non-uniform and is characterized by four well defined peaks, i.e. "hot spots". These peaks correspond to regions in *k* space where there is a small energy gap between the occupied and empty bands: this small gap is a consequence of the spin-orbit coupling within the Mn 3d band.

Moreover, a large AHC of ~900 $(\Omega.cm)^{-1}$ has previously been shown theoretically by considering a slightly different spin configuration (*8*). In the latter case, a non-vanishing component of the AHC can be obtained in the *yz* plane. The reason for this difference is straightforward; in the spin configuration shown in Fig. 1B, if the spins parallel to *y* tilt slightly in the *x* direction, a nonzero Berry curvature in the *yz* plane will result. In all of these configurations the AHE is zero in the *xy* plane if one strictly considers only an in-plane spin structure.

It is well known that like the anomalous Hall effect, the intrinsic spin Hall effect also originates from the band structure. The strong entanglement between occupied and unoccupied states near the Fermi level should give rise to a large contribution to both anomalous and spin Hall conductivities. Motivated by our large AHC, we have also theoretically analyzed the spin Hall conductivity (SHC) and we predict that a large SHC of 1100 (ℏ/e) $(\Omega.cm)^{-1}$ can be obtained in $Mn_3Ge$. The theoretical SHC in $Mn_3Ge$ is comparable to that of Pt (*17*) and larger than the SHC recently reported in $Mn_3Ir$ which also exhibits a triangular AFM structure (*18*). In the present work we experimentally show evidence for an AHC in a hexagonal single crystal of $Mn_3Ge$ that agrees well with our theoretical predictions of AHC.

To probe the AHE in $Mn_3Ge$, we have carried out detailed Hall effect measurements as a function of magnetic field over temperatures varying from 2K to 400 K. Figure 2A shows the Hall resistivity $\rho_H$ versus magnetic field *H* measured for current (*I*) along [0001] and *H* parallel to [01-10]. Most interestingly, $\rho_H$ saturates in modest magnetic fields attaining a large saturation value of 5.1 μΩ cm at 2 K and ~1.8 μΩ cm at 300 K. The Hall conductivity for the same *I* and *H* configuration (that we label configuration I) ($\sigma_{xz}$) calculated from $\sigma_H = -\rho_H/\rho^2$, where ρ is the normal resistivity, exhibits a large value of ~500 $(\Omega.cm)^{-1}$ at 2 K and a modest value of ~50 $(\Omega.cm)^{-1}$ at 300 K (Fig. 2B). Since both $\rho_H$ and $\sigma_H$ saturate in magnetic field these clearly reflect an anomalous Hall effect that is large in magnitude. To study whether the experimental AHE has an anisotropy, as predicted by theory, we measured $\rho_H$ with *I* along [01-10] and *H* parallel [2-1-10] ($\sigma_{yz}$). In this configuration II, $\rho_H$ is slightly smaller compared to configuration I (~4.8 μΩ cm at 2 K and 1.6 μΩ cm at 300 K) (Fig. 2C). Although $\sigma_H$ ($\sigma_{yz}$) is reduced at 2 K in configuration II compared to configuration I (to about 150 $(\Omega.cm)^{-1}$), a similar value of $\sigma_H = 50$ $(\Omega.cm)^{-1}$ at 300 K is foucnd for both configuration I and II (Fig. 2D). In both cases $\rho_H$ is negative (positive) for positive (negative) fields. Finally, in a third configuration (III) we apply *I* along [2-1-10] and *H* parallel [0001]. In this configuration, we observe small values of both $\rho_H$ and $\sigma_H$ ($\sigma_{xy}$) at all temperatures (Fig. 2E,F) and, moreover, the sign of AHE is opposite to that for configurations I and II.

To confirm that the observation of a large AHE in the present system originates from the non-collinear antiferromagnetic (AFM) spin structure, we have performed magnetization measurements with the field parallel to different crystallographic directions. The temperature dependence of magnetization, *M* (*T*), indicates a $T_C$ of 380 K for the present Mn$_3$Ge single crystal (*19*), only slightly lower than that reported in the literature. The *M* (*H*) loop measured with the field parallel to [2-1-10] displays a tiny spontaneous magnetization of 0.005 $\mu_B$/Mn at 2 K (Fig. 3). The zero field moment is slightly larger (0.006 $\mu_B$/Mn at 2 K) when the field is applied parallel to [01-10]. An extremely soft magnetic behavior is found in both cases with coercive fields ($H_C$) of less than 20 Oe. These magnetization values agree well with previous results that show an in-plane magnetic moment of ~0.007 $\mu_B$/Mn (*12*). Although it is expected that the Mn moments lie only in the *a-b* plane, we observe an order of magnitude moment less of only 0.0007 $\mu_B$/Mn at 2 K for field parallel to [0001]. This indicates that there could be a very small tilting of the Mn-spins towards the *c*-axis. The *M* (*H*) loops measured at 300 K for different field orientations are similar to the 2 K data, except that the high-field magnetization strongly decreases for *H* parallel to [0001] (inset of Fig. 3).

From our magnetization data, it is clear that the present Mn$_3$Ge sample possesses a nearly zero net magnetic moment. Hence, we conclude that the extremely large AHE depicted in Fig. 2 must be related to the non-collinear AFM spin structure, as predicted by theory. The observation of large AHC in *xz* and *yz* planes and a very small effect in *xy* plane is also consistent with our theoretical calculations.

Although our calculations show a non-zero AHC only in the *xz* plane this is true for the particular spin configuration considered (see Fig. 1B). By symmetry the *x* and *y* directions must be symmetry related so that there will be equal contributions to the measured AHE for the various symmetry related spin configurations (by rotation by ±120 degree in Fig. 1B) or the spin configurations can be rotated by the applied field due to the small in-plane magnetic moment.

It is interesting to compare the magnitude of the present AHE with that exhibited by ferromagnetic Heusler compounds since the hexagonal $Mn_3Ge$ structure evolves from the cubic Heusler structure by distortion of the lattice along the (111) direction. For example, the ferromagnetic Heusler compounds $Co_2MnSi$ and $Co_2MnGe$, that exhibit a net magnetic moment of ~5 $\mu_B$/Mn (more than 2 orders of magnitude higher by comparison to the present h-$Mn_3Ge$), display a maximum $\rho_H$ ~ 4 $\mu\Omega$ cm (*20*). In the same study it was shown that $\rho_H$ decreases by two orders when the magnetization is reduced to nearly half in $Cu_2MnAl$ (*20*). A similar scaling of $\rho_H$ with magnetization has also been demonstrated by Zeng *et al.* for ferromagnetic $Mn_5Ge_3$ (*2*). Moreover, it has been also well established that the Hall conductivity in many ferromagnetic materials scales with their normal conductivity (*20-22*). In the present case, the Hall conductivity measured with the current along different crystallographic directions follows a similar behavior to that of the normal conductivity (*19*). We also note that similar data for the AHE in the non-collinear AFM $Mn_3Sn$ has been recently published by Nakatsuji *et al.* (*23*). However, $Mn_3Ge$ exhibits distinctions in its magnetism as well as its AHE in comparison to $Mn_3Sn$. In particular, the maximum AHC in the present system is nearly three times higher than that of $Mn_3Sn$. Moreover, $Mn_3Ge$ does not show any secondary transition such as the cluster glass phase observed at low temperatures in $Mn_3Sn$ (*23*).

According to theoretical predictions, the non-vanishing Berry curvature arising from the non-collinear triangular AFM structure is responsible for the observed large AHE. In such a scenario, we would expect to observe an AHE without applying any magnetic field. However, the special spin arrangement in both h-Mn$_3$Ge and h-Mn$_3$Sn can lead to the formation of antiferromagnetic domains. Therefore, in zero magnetic field the direction of the Hall voltage may be different in each AF domain, resulting in a vanishing AHC. The presence of a weak in-plane ferromagnetic moment ensures that a small magnetic field in one direction within the $a-b$ plane is sufficient to eliminate the domains and hence allows the establishment of a large AHE. In particular, neutron diffraction measurements have established that an applied magnetic field does not substantially effect the 120 degree triangular antiferromagnetic spin structure, but the magnetic field does result in a rotation of the spin-triangle opposite to the in-plane field component (*14, 24*). Therefore, switching the direction of an in-plane field from positive to negative results in switching the direction of the staggered moment in the triangular spin structure and consequently causes a change in the sign of the anomalous Hall voltage. The small variation of the anomalous Hall signal at high fields likely originates from the normal component of the Hall signal or the magnetic susceptibility of the magnetic structure.

To establish that the presence of the weak in-plane ferromagnetism in Mn$_3$Ge has virtually no role on the observed AHE, we have carried out angular dependent measurements of the AHE for various field and current configurations. In configuration I shown in Fig. 4A, the current was applied along *z* and the voltage was measured in the *xz* plane, resulting in $V_x$. In this configuration the initial field direction for $\theta$=0 is along *y*. Then the field was rotated in the *y−z* plane. In these conditions, the AHC ($\sigma_{yz}$) remains constant up to $\theta$=90°, where it suddenly changes sign. A similar sign change occurs again at $\theta$=270°. This behavior can be explained as follows; with increasing $\theta$, although the in-plane magnetic field decreases, a small component of the field lies parallel to the *a−b* plane up to $\theta$=90°. This small in-plane field maintains the chirality of the triangular spin structure in a single direction. At angles greater than 90° the in-plane field changes its direction resulting in a change in the chirality of the spin structure which produces the sign change in the AHE. A similar phenomenon happens at 270° (Fig. 4B). In configuration II, we apply the current along *y* with the voltage measured in the *yz* plane ($V_z$), and the field was rotated in the $x-z$ plane. In this case, $H$ is always perpendicular to $I$. The angular dependence of the AHC ($\sigma_{yz}$) follows the same behavior as that of configuration I. In both cases, a large and constant AHC was obtained when at least a small field parallel to the in-plane spin structure is present. We now show that the AHC measured in the $xy$ plane ($\sigma_{xy}$) always remains nearly zero even when there is a nonzero field component parallel to the plane of the triangular spin structure. In configuration III, $V_y$ was measured in the $xy$ plane with $I$ along $x$ (Fig. 4C). The field was rotated in the $z$-$y$ plane. In this configuration, a component of the magnetic field always remains parallel to the $a-b$ plane, except $\theta=0$ and 180°. However, $\sigma_{xy}$ remains nearly zero regardless of the field direction. A small non-vanishing value of $\sigma_{xy}$ may appear from a slight canting of the spins along the $z$ ($c$) direction. The feasibility of a non-planar structure in $Mn_3Ge$ has also been discussed in our previous theoretical work (*8*). From these angle dependent measurements, it can be readily concluded that the small residual in-plane ferromagnetic component as well as the applied magnetic field play no role in the observed large AHE in $Mn_3Ge$.

In summary, we have experimentally demonstrated that the non-collinear antiferromagnet $Mn_3Ge$ exhibits a large anomalous Hall effect, even at room temperature. The origin of this exotic effect is discussed in terms of a non-vanishing Berry curvature in momentum space. Antiferromagnets are of significant interest for spintronic applications (*25-27*), since, in contrast to ferromagnets, they do not produce any undesirable dipole fields. Therefore, the present antiferromagnetic material with a large AHE, which is easily controlled by current, can be exploited for spintronics, without the drawbacks of ferromagnets. Finally, our theoretical finding of a large spin Hall effect in $Mn_3Ge$ will motivate further studies to explore other phenomena derived from the Berry-phase curvature in antiferromagnets.

**Note:** Here we note that during the writing of present manuscript we became aware of similar work on a large anomalous Hall effect in $Mn_3Sn$ by Nakatsuji et al. Ref. (*23*). We have compared the present findings in $Mn_3Ge$ with that of $Mn_3Sn$ in the main text.

**MATERIALS AND METHODS:**

For the single crystal growth, stoichiometric quantities of pure elements were weighed and pre-melted in an alumina crucible in an induction furnace. The ingot was ground and put in an alumina crucible which was sealed in a quartz ampule. The single crystal was grown with the Bridgman-Stockbarger-technique. The growth temperature was controlled at the bottom of the ampule. The material was heated up to 980°C, remained there for one hour to ensure homogeneity of the melt, and then was cooled down slowly to 740°C. The sample was annealed at 740°C for 15 hours followed by quenching to room temperature in argon gas. The sample was cut into 5 slices perpendicular to the growth direction. The dimensions of each slice were approximately 6 mm in diameter and 2 mm in thickness. Single crystal X-ray diffraction was carried out on a Bruker D8 VENTURE X-ray diffractometer using Mo-K radiation and a bent graphite monochromator. The magnetization measurements were carried out on a vibrating sample magnetometer (MPMS 3, Quantum Design). The transport measurements were performed with low-frequency alternating current (ACT option, PPMS 9, Quantum Design). Our ab-initio DFT calculations were performed by the Vienna Ab-initio Simulation Package (VASP) with projected augmented wave (PAW) potential (*28*). The exchange and correlation energy was considered in the generalized gradient approximation (GGA) level (*29*). Further details of the calculations are given in the Supplementary Information.

**Acknowledgements:** We thank Christoph Geibel for helpful discussions and use of the Bridgman furnace. This work was financially supported by the Deutsche Forschungsgemeinschaft DFG (Projects No.TP 1.2-A and No. 2.3-A of Research Unit FOR 1464 ASPIMATT) and by the ERC Advanced Grant No. (291472) "Idea Heusler". **Author Contributions**: AKN performed the transport and magnetic measurements with help of CS, NK and WS. YS, BY and JK carried out the theoretical calculation. JEF synthesized the single crystal. JEF, AK and J Karel performed the structural characterization. AKN and SSPP wrote the manuscript with substantial contributions from all authors. CF and SSPP inspired and jointly supervised the project. **Author Information**: The authors declare no competing financial interests. Correspondence and requests for materials should be addressed to AKN (nayak@cpfs.mpg.de) and SSPP (stuart.parkin@mpi-halle.mpg.de).

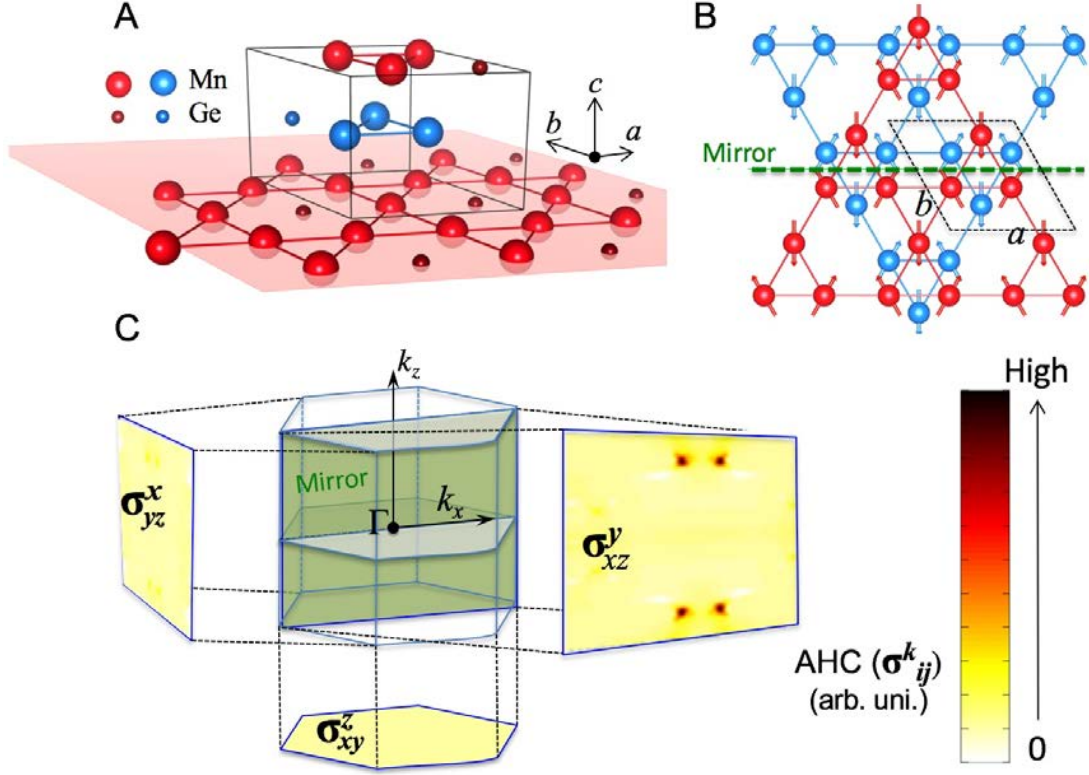

Fig 1. **Crystal structure, magnetic structure and Berry curvature of $Mn_3Ge$**. **A**, Crystal structure showing the two layers of Mn-Ge atoms stacked along the $c(z)$ axis. Red and blue spheres represent atoms lying in the $z = 0$ and $z = c/2$ planes, respectively. Large and small spheres represent Mn and Ge atoms, respectively. In each layer, the Mn atoms form a Kagome-type lattice. **B**, The 120 degree AFM configuration in the $z = 0$ and $z = c/2$ planes, respectively. Only the Mn atoms are shown with their moments represented by arrows. The green dashed line indicates the mirror plane, i.e. the $xz$ plane. Corresponding mirror reflection plus a translation by $c/2$ along the $z$–axis transforms the system back into itself, by exchanging two magnetic planes. **C**, The first Brillouin zone and the momentum-dependent AHC. The mirror plane, i.e. $k_x$–$k_z$ plane, is highlighted in green. The component of the AHC tensor ($\sigma^k_{ij}$) is shown in the $k_i$–$k_j$ plane, which corresponds to the Berry curvature of all occupied electronic states integrated over the $k_k$ direction. Here only the $\sigma^y_{xz}$ component exhibits large values.

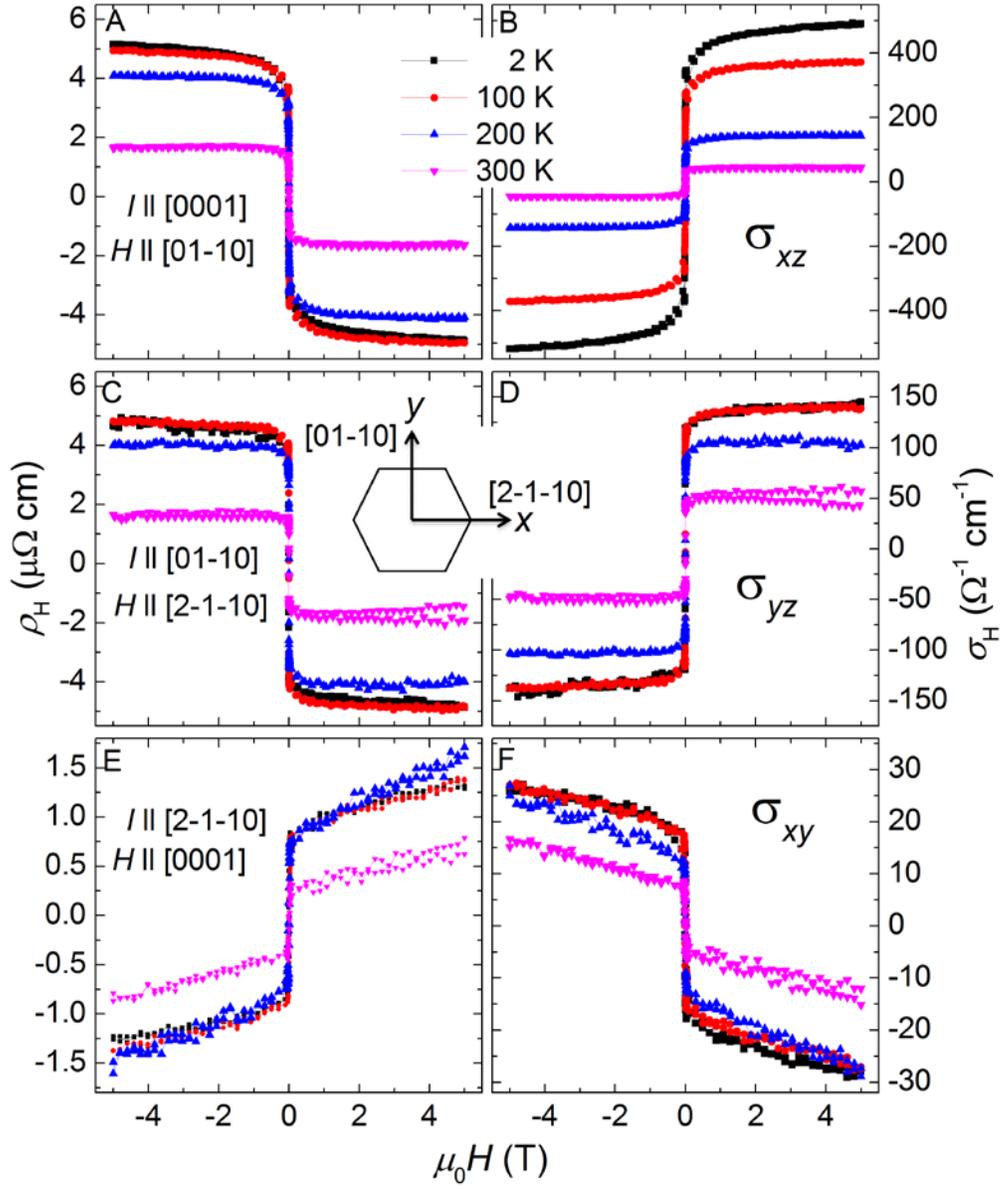

Fig. 2. **Anomalous Hall effect. A,C,E**, Hall resistivity ($\rho_H$) and **B,D,F**, Hall conductivity ($\sigma_H$) as a function of magnetic field (*H*), measured at four representative temperatures between 2 and 300 K, for three different current and magnetic field configurations. The hexagon in the inset to **C,D** shows the two in-plane directions used for the current and magnetic field in the present measurements.

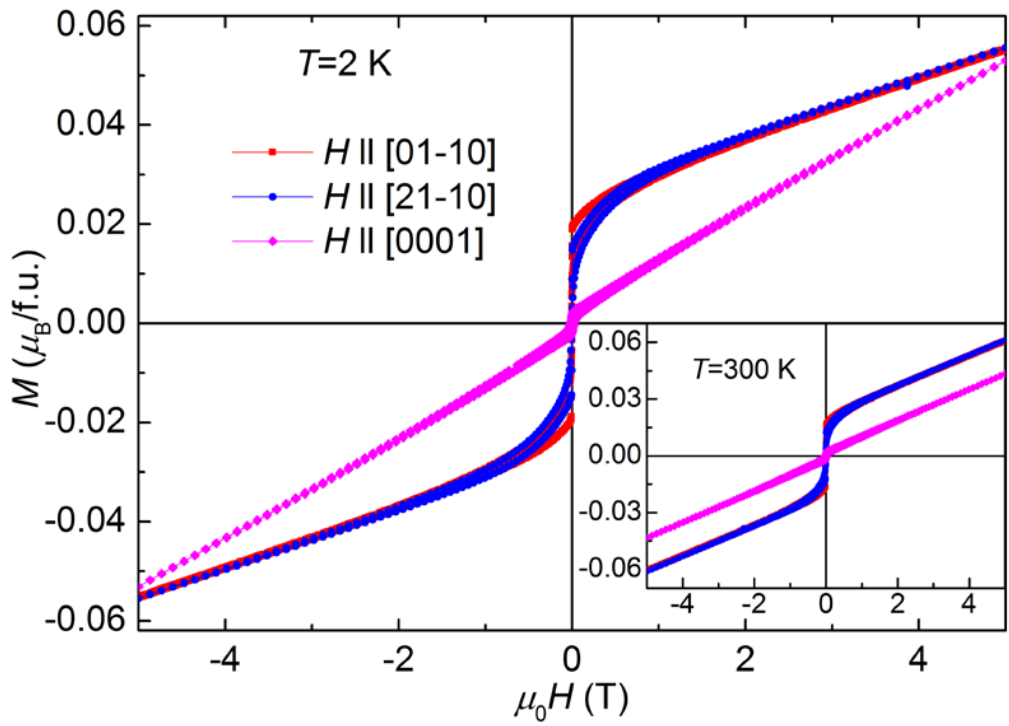

Fig. 3. **Magnetic properties**. Field dependence of magnetization, $M$ ($H$), measured at 2 K, with field parallel to the three orientations used in the AHE measurements. The inset shows $M$ ($H$) loops measured at 300 K.

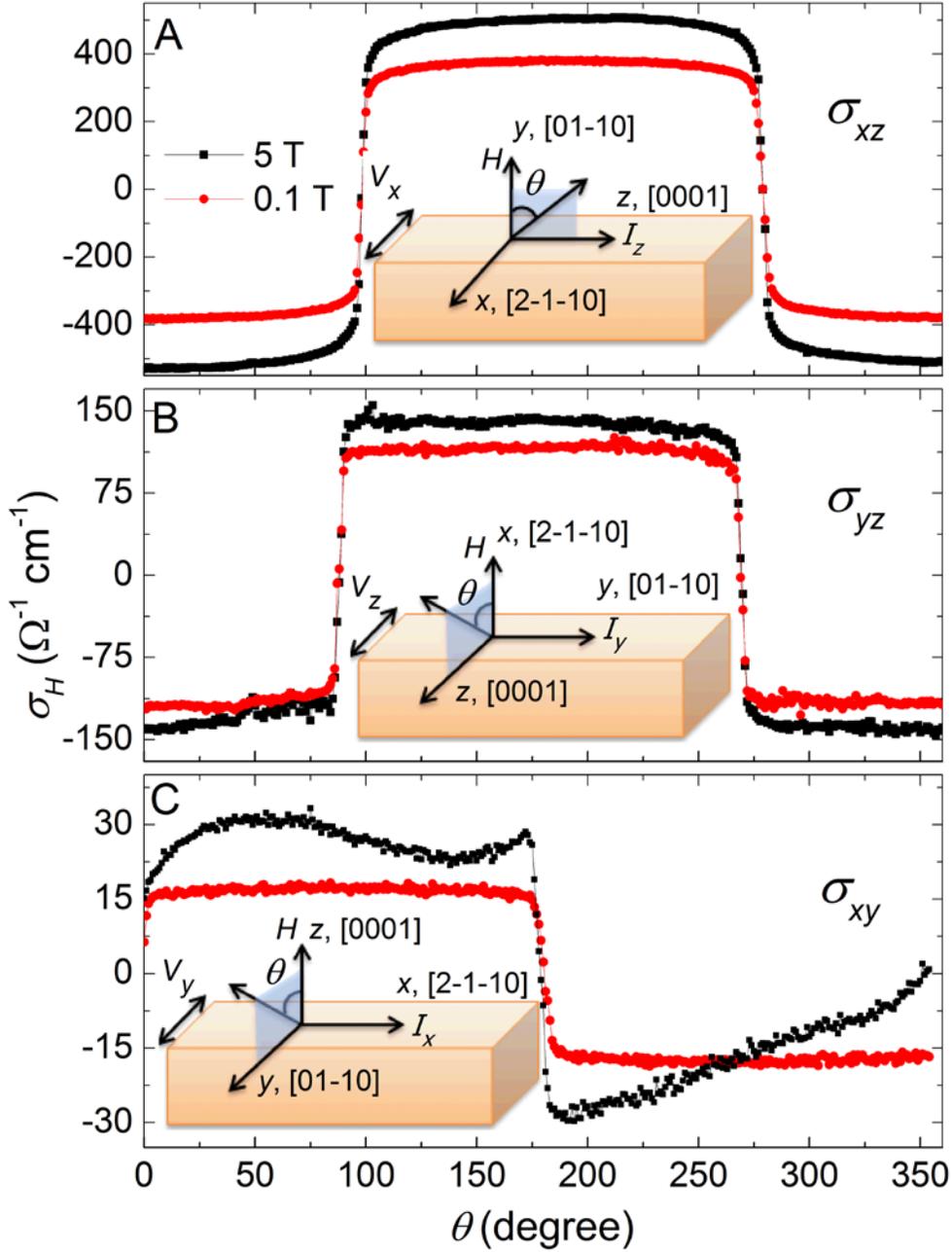

Fig. 4. **Angular dependence of anomalous Hall effect**. Angular ($\theta$) dependence of the anomalous Hall conductivity measured for the three current and field configurations used in Fig. 2. A schematic diagram of the sample geometry is shown for each configuration. The rotation was performed along **A**, $y-z$ for $I$ along $z$ [0001], **B**, $x-z$ for $I$ along $y$ [01-10] and **C**, $y-z$ for $I$ along $x$ [2-1-10]. $\theta=0$ corresponds to $H$ oriented along $y$, $x$ and $z$ in **A, B** and **C**, respectively. We have used the same $x$, $y$, $z$ coordinates as shown in the theoretical spin arrangement in Fig. 1**B**.

# Supplementary Information:

Our ab-initio DFT calculations were performed by the Vienna Ab-initio Simulation Package (VASP) with projected augmented wave (PAW) potential (*28*). The exchange and correlation energy was considered in the generalized gradient approximation (GGA) level (*29*). The experimental lattice structure was used in the calculations. We projected the Bloch wave functions into Wannier functions as s, p and d atomic-like orbitals (*30*) and derived the tight-binding Hamiltonian for Mn$_3$Ge. The intrinsic transverse Hall conductivity can be understood as the integral of the Berry curvature for the occupied states in the reciprocal space (*1*). Hence, we have calculated the Berry curvature from the ab-initio tight binding model, to interpret the measured anomalous Hall effect,

The intrinsic anomalous Hall conductivity (AHC) tensor, $\sigma_{\alpha\beta}^{\gamma}$, is calculated as the sum of the Berry curvature for a given electronic state, n, that is defined by the quantity, $\Omega_{\alpha\beta}^{n,\gamma}(\vec{k})$, over all the occupied electronic states, as described by the equation:

$$\sigma_{\alpha\beta}^{\gamma} = ie^2\hbar \; \Sigma_n \int_{\vec{k}} \frac{d\vec{k}}{(2\pi)^3} f(E_{n,\vec{k}}) \; \Omega_{\alpha\beta}^{n,\gamma}(\vec{k}) \tag{1}$$

where α, β, γ can be any of x,y,z, $f(E_{n,\vec{k}})$ is the Fermi-Dirac distribution, and $\Omega_{\alpha\beta}^{n,\gamma}(\vec{k})$, a is given by

$$\Omega_{\alpha\beta}^{n,\gamma}(\vec{k}) = Im \; \Sigma_{n'\neq n} \frac{<u(n,\vec{k})|\hat{v}_{\alpha}|u(n',\vec{k})><u(n',\vec{k})|\hat{v}_{\beta}|u(n,\vec{k})> - (\alpha \leftrightarrow \beta)}{\left(E_{n,\vec{k}} - E_{n',\vec{k}}\right)^2} \tag{2}$$

where, $E_{n,\vec{k}}$ is the eigenvalue for the *n-th* eigen state $|u(n,\vec{k})>$ at the $\vec{k}$ point, and $\hat{v}_{\alpha} = \frac{1}{\hbar}\frac{\partial H(\vec{k})}{\partial k_{\alpha}}$ is the velocity operator. A *k*-grid of 101 × 101 × 101 was used in the integral over the Brillouin zone.

We have calculated the intrinsic spin Hall conductivity $\sigma_{xy}^{z}$ with field along *y*, spin current along *x*, and spin polarization in *z* direction. The spin Hall conductivity was calculated by using the Kubo formula approach (*31*),

$$\sigma_{xy}^{z} = \frac{e\hbar}{V}\sum_{\vec{k}}\sum_{n} f_{\vec{k},n}\left(-2\operatorname{Im}\sum_{n'\neq n}\frac{<u_{n\vec{k}}|\hat{J}_{x}^{z}|u_{n'\vec{k}}><u_{n'\vec{k}}|\hat{v}_{y}|u_{n\vec{k}}>}{\left(E_{n\vec{k}}-E_{n'\vec{k}}\right)^{2}}\right)$$

where the spin current operator $\hat{J}_{x}^{z} = \frac{1}{2}\{\hat{v}_{x}\quad \hat{s}_{z}\}$, the spin operator $\vec{s}_{z}$.

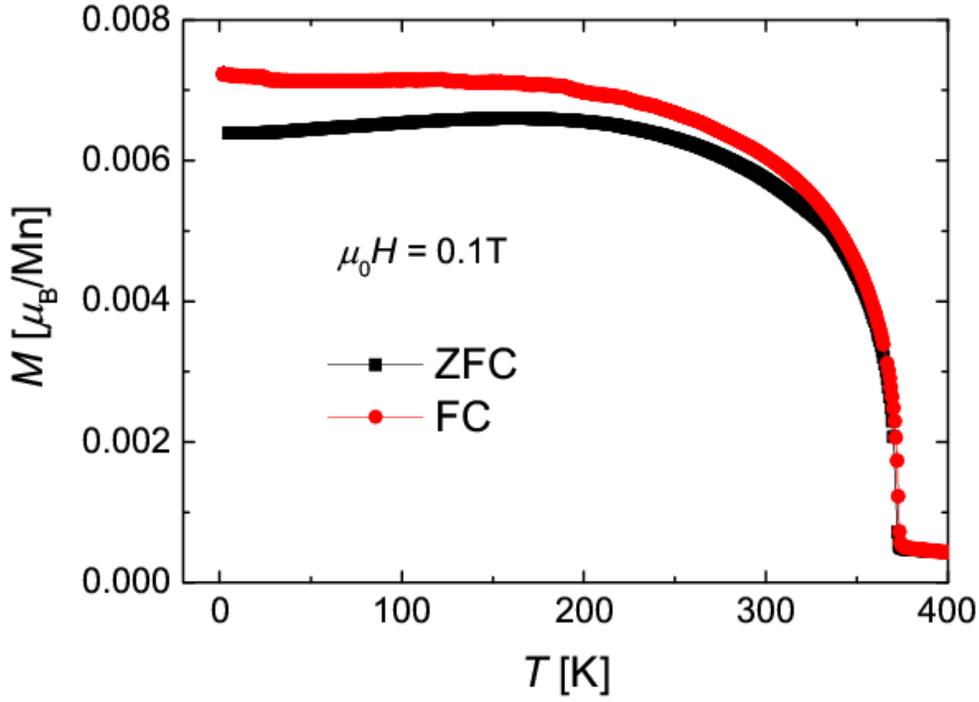

**Fig. S1. Temperature dependence of magnetization**. $M$ ($T$) measured with field parallel to a-b plane.

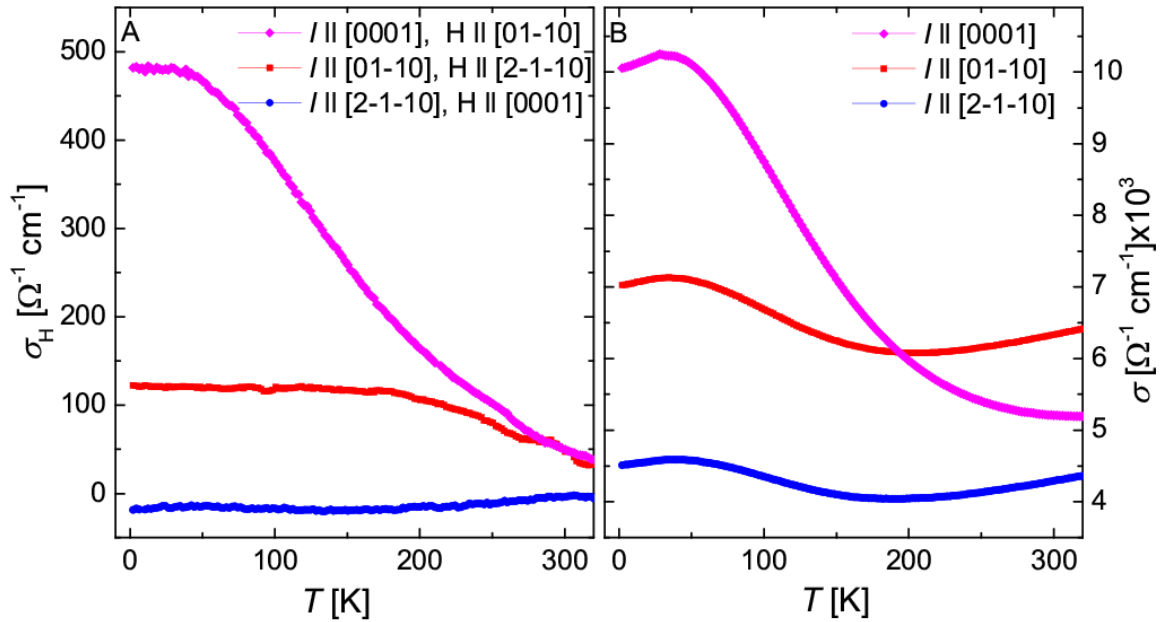

**Fig. S2. Conductivity with temperature**. **A**, Temperature dependence of Hall conductivity measured with the current along three different crystallographic directions. The magnetic field was kept at 0.1 T for all measurements. **B**, Temperature dependence of the normal conductivity measured in zero magnetic field for the same current configurations.

The temperature variation of the anomalous Hall conductivity measured in different current directions is shown in Fig. S2A. Since Mn$_3$Ge exhibits extremely low coercivity, all measurements have been performed in the presence of a small field of 0.1 T. The Hall conductivity measured with *I* along [0001] and *H* along [01-10] exhibits a maximum value of 480 ($\Omega$cm)$^{-1}$ at 2 K and remains constant up to 40 K. Above 40 K, it decreases almost linearly with temperature leading to a value of approximately 40 ($\Omega$cm)$^{-1}$ at room temperature. By contrast, the temperature dependence of the low field Hall conductivity for *I* along [01-10] and *H* along [2-1-10] virtually remains constant with temperature, showing only a small decrease above 200 K. Similarly, Hall conductivity measured with *I* along [2-1-10] and H along [0001], which exhibits a negative value, remains nearly unchanged with temperature. To elucidate the relationship between hall conductivity and normal conductivity in Mn$_3$Ge, we have plotted the temperature dependence of the normal conductivity measured for different current profiles in Fig. S2B. Indeed, we found that for same current direction the normal conductivity follows a similar temperature dependent behavior as that of Hall conductivity.